\documentclass[9pt,conference]{IEEEtran}
\IEEEoverridecommandlockouts

\usepackage{blkarray}                                      
\usepackage{algpseudocode}                                 
\usepackage[linesnumbered,ruled,vlined]{algorithm2e}
\usepackage{graphicx}                                      
\usepackage{amsmath}
\usepackage{amssymb}
\usepackage{amsfonts}
\usepackage{amsthm}
\usepackage[mathcal]{eucal}
\usepackage{mathrsfs}
\usepackage{booktabs}
\usepackage{enumerate}
\usepackage{multirow}
\usepackage[subrefformat=parens,farskip=0pt,justification=centering]{subfig}
\usepackage{color}
\usepackage{cite}                                          
\usepackage{comment}                                       
\usepackage{soul}                                          
\soulregister\cite7
\soulregister\ref7
\soulregister\pageref7
\usepackage{etoolbox}                                      
\usepackage{url}
\usepackage{nth}                                           
\usepackage{bm}                                            
\usepackage{courier}
\usepackage{balance}
\usepackage{threeparttable}
\usepackage{xcolor,colortbl}
\usepackage{footnote}
\usepackage{listings}
\usepackage{setspace}                                      

\usepackage{verbatim}
\usepackage[bookmarks=false]{hyperref}
\hypersetup{
    colorlinks = true,
    citecolor  = blue,
    linkcolor  = blue,
    urlcolor   = blue,
}
\usepackage{tikz}
\usetikzlibrary{patterns,snakes}
\usetikzlibrary{positioning,calc,fit,decorations.pathmorphing,shapes.geometric, shapes.gates.logic.US, calc}
\usetikzlibrary{arrows,arrows.meta,decorations.markings,shapes,shapes.arrows}
\usetikzlibrary{decorations,decorations.pathreplacing}
\usetikzlibrary{backgrounds}
\usepackage{filecontents}                                  
\usepackage{pgfplots}
\usepackage{pgfplotstable}
\usepackage{scalefnt}
\pgfplotsset{compat=newest}
\usepackage{caption}
\usepackage{cleveref}
\Crefformat{figure}{Fig.~#2#1#3}                           
\Crefname{subfigure}{Fig.}{Figs.}
\Crefname{figure}{Fig.}{Figs.}
\Crefformat{table}{TABLE~#2#1#3}                           
\usepackage[figuresright]{rotating}

\definecolor{CUHKorange}{RGB}{244,106,18} 
\definecolor{CUHKblue}{RGB}{0,111,190}    
\definecolor{CUHKgreen}{RGB}{0,127,128}   
\definecolor{CUHKred}{RGB}{228,46,36}     
\definecolor{CUHKyellow}{RGB}{198,148,34} 
\definecolor{CUHKdark}{RGB}{114,44,114}   
\definecolor{CUHKmiddle}{RGB}{144,44,144} 
\definecolor{CUHKlight}{RGB}{167,44,167} 
\definecolor{CUHKpurple}{RGB}{117,15,109}
\definecolor{CUHKgold}{RGB}{221,163,0}
\definecolor{CUHKribbon}{RGB}{244,223,176}
\definecolor{CUHKblack}{RGB}{34,24,21}

\renewcommand{\bf}[1]{\textbf{#1}}
\renewcommand{\tt}[1]{\texttt{#1}}

\newcommand\blfootnote[1]{                 
  \begingroup
  \renewcommand\thefootnote{}\footnote{#1}%
  \addtocounter{footnote}{-1}%
  \endgroup
}

\usepackage{tcolorbox}
\tcbuselibrary{skins,breakable}
    {\endtcolorbox}

\newtheorem{myproblem}{\textbf{Problem}}

\crefname{mytheorem}{Theorem}{Theorems}
\crefname{mylemma}{Lemma}{Lemmas}
\crefname{myclaim}{Claim}{Claims}
\crefname{myproperty}{Property}{Properties}
\crefname{mycorollary}{Corollary}{Corollaries}

\algrenewcommand\textproc{\texttt}

\makeatletter
\let\OldStatex\Statex
\renewcommand{\Statex}[1][3]{%
  \setlength\@tempdima{\algorithmicindent}%
  \OldStatex\hskip\dimexpr#1\@tempdima\relax
}
\makeatother

\RequirePackage[normalem]{ulem} 
\RequirePackage{color}\definecolor{RED}{rgb}{1,0,0}\definecolor{BLUE}{rgb}{0,0,1} 

\begin{document}

\title{\huge
A Systematic Approach for Multi-objective Double-side Clock Tree Synthesis}

\author{
Xun Jiang$^{1}$, Haoran Lu$^{1}$, Yuxuan Zhao$^{2}$, Jiarui Wang$^{1,3}$, Zizheng Guo$^{1}$, Heng Wu$^{1}$, Bei Yu$^{2}$, Sung Kyu Lim$^{4}$, \\Runsheng Wang$^{1,5,6}$, Ru Huang$^{1,5,6}$ and Yibo Lin$^{1,5,6*}$\\
  {\large{$^{1}$School of IC, Peking University}}\ 
  {\large{$^{2}$Department of CSE, The Chinese University of Hong Kong}}\\
  {\large{$^{3}$School of CS, Peking University}}\ 
  {\large{$^{4}$School of ECE, Georgia Institute of Technology}}\\
  {\large{$^{5}$Institute of EDA, Peking University, Wuxi}}\
  {\large{$^{6}$Beijing Advanced Innovation Center for Integrated Circuits}}\\
}

\maketitle
\vspace*{-1.4cm}
\blfootnote{\rule{0.27\linewidth}{0.1pt} \\
$^*$Corresponding author: Yibo Lin (yibolin@pku.edu.cn)}

\begin{abstract}
As the scaling of semiconductor devices nears its limits, utilizing the back-side space of silicon has emerged as a new trend for future integrated circuits. With intense interest, several works have hacked existing backend tools to explore the potential of synthesizing double-side clock trees via nano Through-Silicon-Vias (nTSVs). However, these works lack a systematic perspective on design resource allocation and multi-objective optimization. We propose a systematic approach to design clock trees with double-side metal layers, including hierarchical clock routing, concurrent buffers and nTSVs insertion, and skew refinement. Compared with the state-of-the-art (SOTA) methods, the widely-used open-source tool, our algorithm outperforms them in latency, skew, wirelength, and the number of buffers and nTSVs.
\end{abstract}

\section{Introduction}
\label{sec:Introduction}

Back-side interconnection \cite{chen2021design,veloso2023backside} has emerged to continue the scaling of semiconductor technologies. With increasingly congested designs and tight timing budgets on the front-side (FS), both the academia \cite{prasad2019buried, lin2024effective, hsu2024bounding,veloso2023backside,bethur2024gnn,bethur2023back,aly2018n3xt,lu2024first} and industry \cite{shamanna2023core,tsmca16} have started to consider utilizing back-side (BS) resources for routing wires, including signal, power, and clock net. Based on the evaluation in \cite{veloso2023backside}, the latency of the clock tree is decreased from 50$ps$ to 20$ps$ with back-side metal layers. However, lacking a systematic double-side clock tree synthesis (CTS) algorithm is impeding further exploration of the potential of back-side resources, considering the new complex design space unfolding before researchers.

\begin{figure}[tb]
    \centering
    \vspace*{-0.6cm}
    \includegraphics[width=0.82\linewidth]{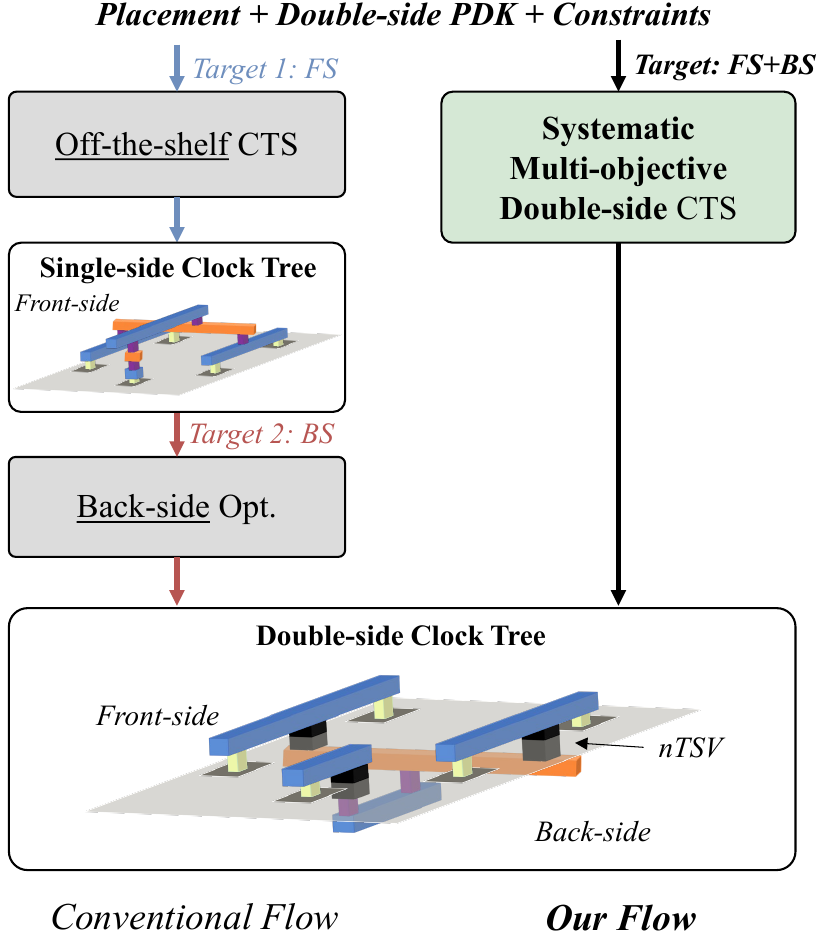}
    \caption{Clock tree synthesis with double-side metal layers.}
    \label{fig:intro-flow}
    \vspace*{-0.5cm}
\end{figure}

As shown in the bottom of \Cref{fig:intro-flow}, the double-side clock net is jointly implemented with connecting the back-side and front-side metal layers via additional nTSVs. Although timing benefits have been reported, the overhead of nTSVs, buffers, and clock wirelength cannot be neglected, as the primary objectives of CTS has included latency, skew, and resource consumption. For instance, many methods, e.g., clock routing \cite{wang2020zero, boese1992zero, edahiro1993clustering,li2024toward}, buffer insertion \cite{van1990buffer, han2018optimal, guthaus2010non, chen2010clock}, useful skew \cite{xi1996useful, shen2010useful, uysal2019latency}, Flip-Flop (FF) clustering \cite{deng2015register, mehta1997clustering, bang2015clock} and 3D clock tree \cite{kim2011clock, yang2011robust, pentapati2019tier} have all once referred to as timing and power optimization techniques involving clock trees. However, considering the challenges of more complex design resource allocation and multi-objective optimization, the exploration of double-side CTS is still insufficient to keep pace with advanced technology.


Existing works \cite{veloso2023backside, bethur2024gnn, vanna2024back, bethur2023back} investigating double-side CTS all follow the incremental flow shown in the left of \Cref{fig:intro-flow}. To elaborate, Veloso et al.~\cite{veloso2023backside} flip the high-level part of a clock tree to the back-side to reduce clock latency to the maximum extent.
Bethur et al.~\cite{bethur2023back} propose leveraging the fanout of driven sinks as the criteria to decide whether the net should be flipped to back-side.
Bethur et al.~\cite{bethur2024gnn} utilize the Graph Neural Network to select the subset of FFs with poor timing performance and flip the connected nets to the back-side.
Vanna-iampikul et al.~\cite{vanna2024back} incorporate the back-side design methodologies of Power Delivery Network (PDN) based on the clock synthesis method by Bethur et al.~\cite{bethur2024gnn}.
Although the benefits of back-side metal layers to CTS are shown explicitly, the incremental flow cannot fully uncover their potential. For instance, the pre-generated single-side clock tree by the off-the-shelf CTS tool only considers the timing model based on the front-side technology parameters to guide the buffer insertion process, whose buffering solutions will deviate from the best one. The follow-up back-side optimization method has to obey these results to assign nets to back side by inserting nTSV, which can severely harm the eventual solution quality due to limited solution space. Thus, a unified design space of buffer and nTSV insertion with efficient solution exploration is urgent for double-side CTS, which is also the core of our work.


In this work, we aim at unleashing the potential of the back-side technology by handling the challenges from double-side CTS systematically. The major contributions of this work are as follows:
\begin{itemize}
    \item We propose a systematic double-side clock tree synthesis framework aiming at pushing the boundary of the cutting-edge back-side technology exploration by multi-objective optimization.
    \item We propose an efficient hierarchical clock routing to reduce clock wirelength and preserve balanced structure.
    \item We propose a concurrent buffer and nTSV insertion based on multi-objective dynamic programming (DP) and an efficient resource-aware end-point buffer insertion as post-processing method for latency, skew, and resource usage optimization.
    \item We propose a novel methodology of design space exploration (DSE) of double-side CTS by our framework.
\end{itemize}

Experimental results demonstrate our superiority over recent works \cite{veloso2023backside, bethur2024gnn, vanna2024back, bethur2023back}. Take the method \cite{veloso2023backside} with extreme optimization on latency as an example, we can optimize the clock latency by 2.223$\times$, skew by 2.464$\times$, number of buffers by 1.010$\times$, clock wirelength by 1.249$\times$, and number of nTSVs by 1.441$\times$, respectively, with 6.922$\times$ speed-up.

The rest of this paper is organized as follows.
In Section \ref{sec:Preliminary}, we demonstrate the preliminaries of our work.
In Section \ref{sec:Algorithm}, we explain the details of our algorithms.
In Section \ref{sec:Results},  we present the results of our work.
In Section \ref{sec:Conclusion}, we conclude our work and discuss further works.

\section{Preliminaries}
\label{sec:Preliminary}


\subsection{Double-side Clock Tree Structure}
\begin{figure}[tb]
    \centering
    \subfloat[Initial Buffered Clock Tree]{\includegraphics[width=0.48\columnwidth]{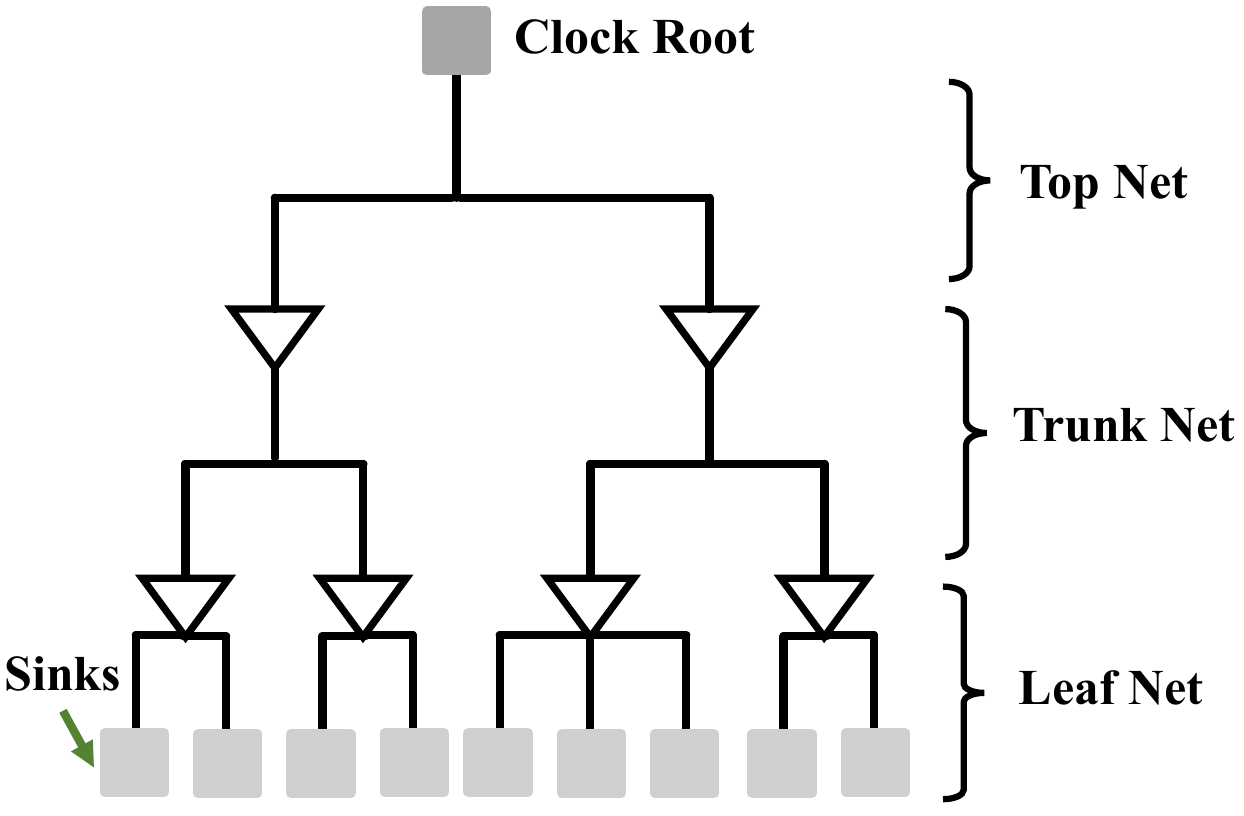}\label{fig:clock-tree1}} \hspace{0.2cm}
    \subfloat[Latency-driven Insertion \cite{veloso2023backside}.]{\includegraphics[width=0.47\columnwidth]{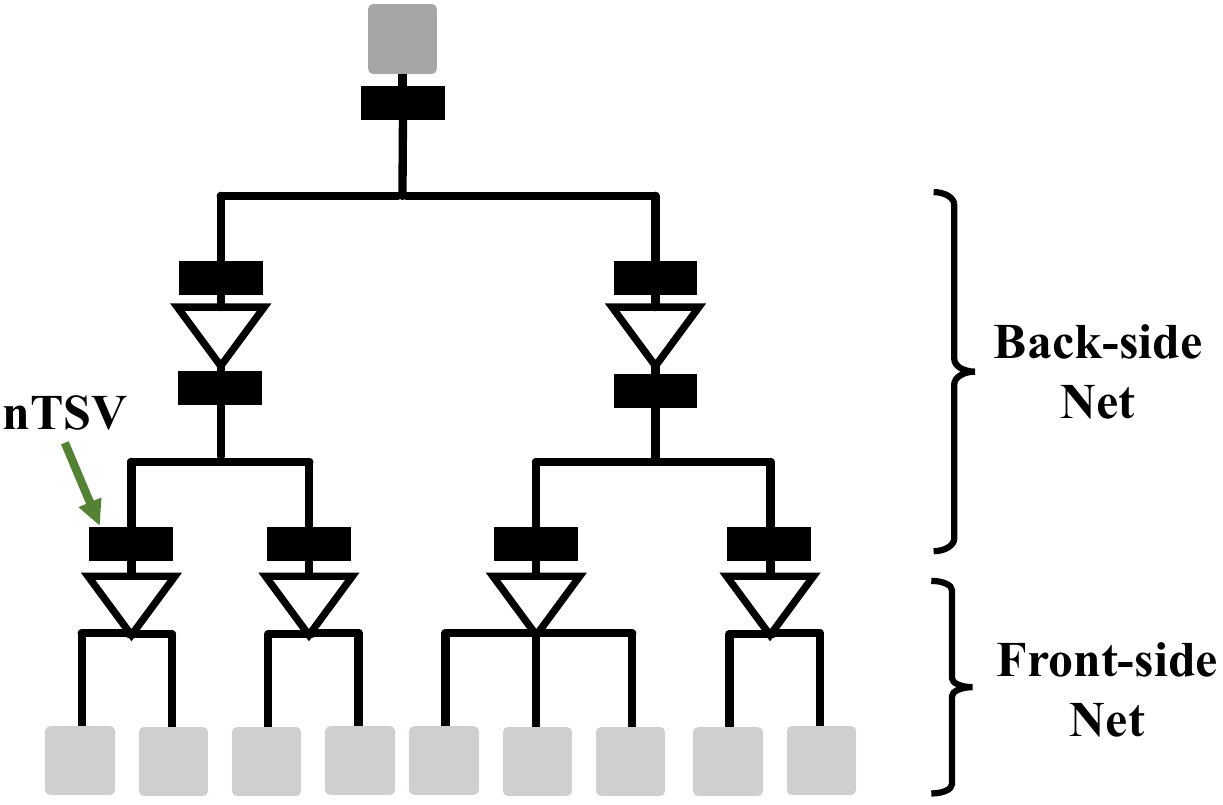}\label{fig:clock-tree2}} \hfill
    \subfloat[Fanout (FO)-driven Insertion \cite{bethur2023back}.]{\includegraphics[width=0.452\columnwidth]{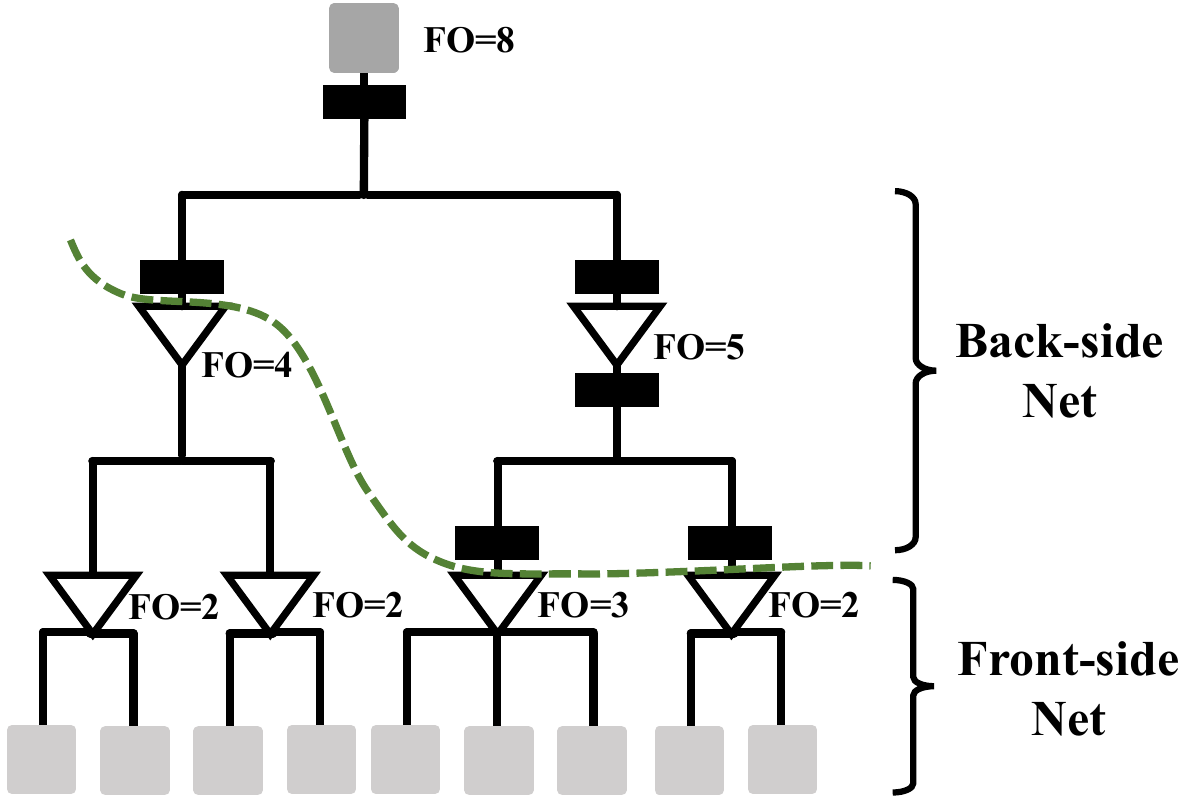}\label{fig:clock-tree3}} \hspace{0.2cm}
    \subfloat[Critical Timing-driven Insertion \cite{bethur2024gnn}.]{\includegraphics[width=0.47\columnwidth]{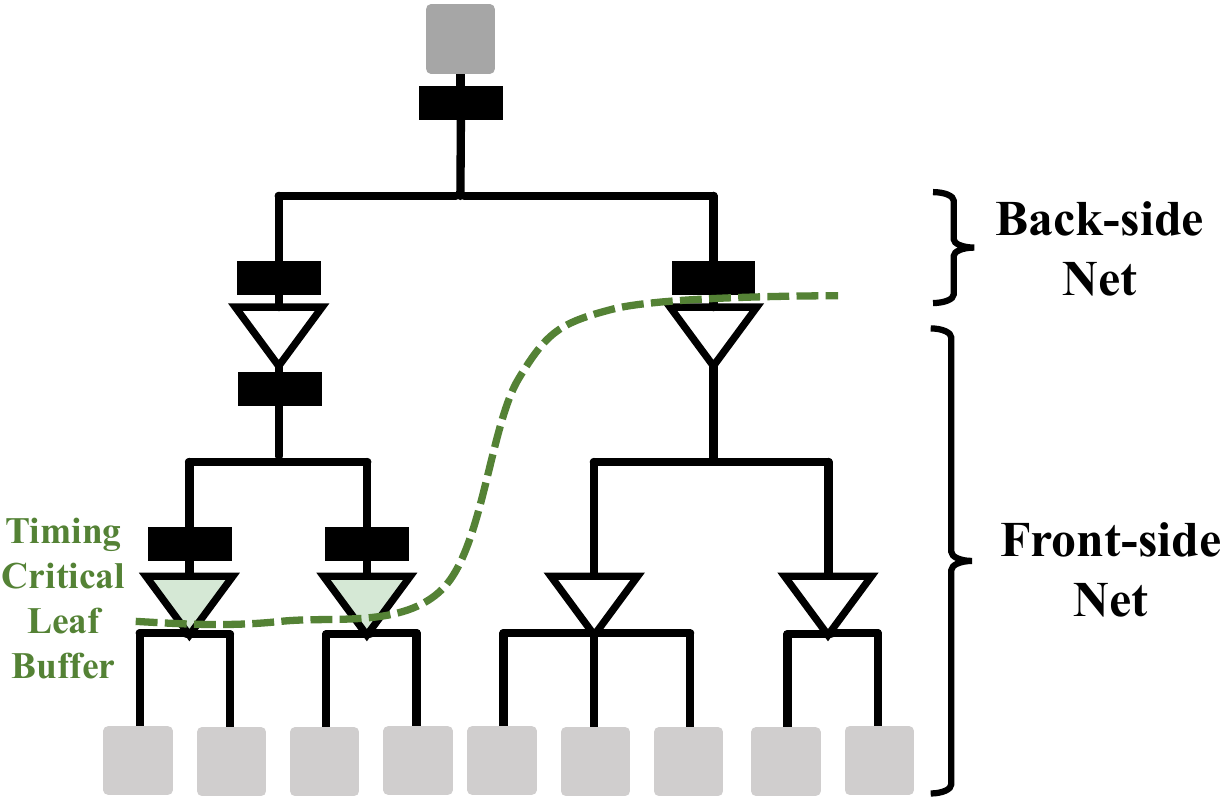}\label{fig:clock-tree4}} \hfill
    \caption{Comparison between the buffered clock tree and the double-side clock trees by different methods to assign top and trunk nets to the back side.}
    \label{fig:traditional-backside}
    \vspace{-0.5cm}
\end{figure}

The comparison of structure between initial buffered  clock tree and double-side clock trees are drawn in \Cref{fig:traditional-backside}. In \Cref{fig:clock-tree1}, the initial buffered  clock tree assign all leaf nets and trunk nets are on the front side. The leaf net includes clock sink pins, while the trunk net encompasses all other nets excluding leaf nets. Top nets, defined by designers as highest-level trunk nets, can be distinguished from trunk nets for clarity. 

We demonstrate three recent post-CTS methods to implement double-side clock tree in \Cref{fig:clock-tree2}, \Cref{fig:clock-tree3}, and \Cref{fig:clock-tree4}, respectively. The post-CTS method from \cite{veloso2023backside} assigns the trunk-level nets to the back-side by inserting nTSVs in \Cref{fig:clock-tree2}. Owing to the input and output pins of the buffer on the front side, multiple nTSVs are incorporated to maintain connectivity between the front and back sides. The delay from source to sink pins can be reduced by the lower unit resistance and capacitance of the back-side metal layers, as most paths traverse these common trunk-level nets. It is noteworthy that extra nTSVs also increase the resource usage which needs to be carefully considered for the design of the entire chip. Therefore, \cite{bethur2023back} utilizes the fanout of driven sinks as the criteria to decide whether the nets should be flipped and \cite{bethur2024gnn} leverages the timing criticality of leaf driving buffers to decide the back-side nets assignment. In a word, these methods try to trade-off the timing benefits and the nTSV utilization but limited to the methodology of separated buffers and nTSVs insertion.

\subsection{Delay Model for Buffers and nTSVs}
\label{sec:prelim-delay}
In the clock tree synthesis with double-side metal layers, buffers and nTSVs are jointly utilized to minimize the latency of the clock tree. We follow previous work \cite{veloso2023backside, bethur2023back, bethur2024gnn} to use the classic \textit{L-type} Elmore delay \cite{hu2015tau} to compute the delay of wires. We set the front-side and back-side metal layer unit capacitance to $c_f$ and $c_b$ and set the front-side and back-side metal layer unit resistance to $r_f$ and $r_b$. Furthermore, we take a wire segment with the length $L$ and set the output driven capacitance of wire segment to $C_d$. 

\begin{figure}[tb]
    \centering
    \subfloat[Inserting Buffer]{\includegraphics[width=0.48\columnwidth]{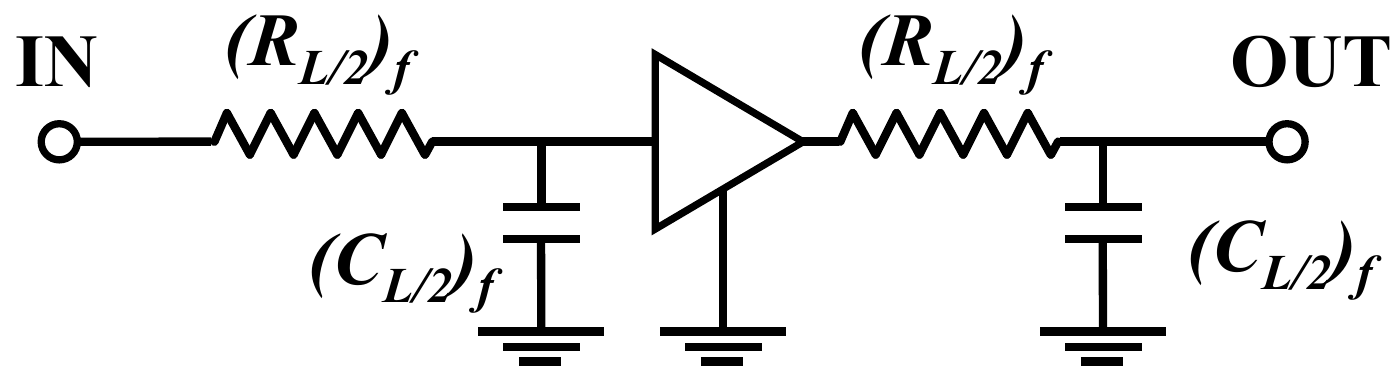}\label{fig:delay-min4}} \hfill
    \subfloat[Inserting nTSV]{\includegraphics[width=0.48\columnwidth]{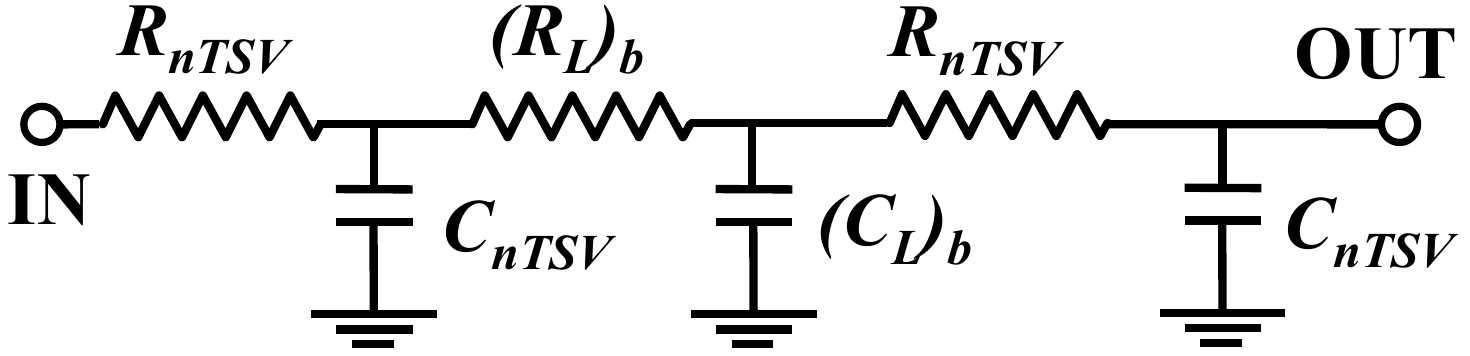}\label{fig:delay-min6}} \hfill
    \caption{Delay modeling for buffer and nTSV insertion.}
    \label{fig:delay-min}
    \vspace{-0.5cm}
\end{figure}


As buffer insertion on the front side in \Cref{fig:delay-min4}, we note $C_b$ as the input capacitance of buffer and $D_{buf}$ as the buffer delay. The delay of inserting a buffer at the middle of the wire is denoted as $D_{bufOn}$, which is computed as follows.
\begin{equation}
    \begin{split}
        D_{bufOn} = & r_f\frac{L}{2}  (c_f\frac{L}{2} + C_b) + D_{buf} + r_f\frac{L}{2}(c_f\frac{L}{2}+C_d)\\
             =& \frac{r_fc_f}{2}L^2 + \frac{r_f(C_b+C_d)}{2}L + D_{buf}.\\
    \end{split}
    \label{eq:buffer-delay}
\end{equation}

We use $C_{nTSV}$ and $R_{nTSV}$ to represent the capacitance and resistance of one nTSV. The delay of inserting two nTSVs at the endpoints of the wire segment in \Cref{fig:delay-min6} is denoted as $D_{nTSVOn}$, which is computed as follows.
\begin{equation}
    \begin{split}
        D_{nTSVOn} =& R_{nTSV}(C_{nTSV}+C_d) + r_bL  (c_bL + C_{nTSV} + C_d)\\ 
        &+ R_{nTSV}(2C_{nTSV}+c_bL+C_d)\\
        =& (r_bc_b)L^2 + (r_bC_{nTSV} + r_bC_d + R_{nTSV}c_b)L\\
        &+ R_{nTSV}(3C_{nTSV}+2C_d).\\ 
    \end{split}
    \label{eq:ntsv-delay}
\end{equation}

In the double-side scenarios, $r_bc_b \ll r_fc_f$ reduce the delay of back-side metal layers. Meanwhile, the buffer can shield the output capacitance to reduce delay and meet maximum driven-capacitance constraint, whereas nTSV cannot. Therefore, the collaborative optimization of buffers and nTSVs insertion is crucial in the double-side CTS.

\subsection{Multi-objective Optimization}
Multi-objective optimization refers to solving problems with multiple conflicting objectives by exploring a set of optimal trade-off solutions. The Pareto frontier \cite{marler2009multi} is the core concept to represent these solutions, where no objective can be improved without degrading another. For the multi-objective optimization, it is essential to leverage Pareto frontier to comprehensively evaluate the performance of algorithms without being influenced by some parameter preferences. The diversity of solutions across the objective space is also important to avoid the algorithm getting stuck in local optimality.

\subsection{Problem Formulation}

\begin{myproblem}[Double-side clock tree synthesis]
    Given the clock net, double-side metal layers, and candidate cells, e.g., nTSV and buffer, construct a double-side clock tree to optimize multiple objectives, e.g., latency, skew, wirelength, buffer count, and nTSV count, under connectivity and electricity constraints. The decision variables encompass the positions of inserted nTSVs and buffers, as well as their mutual topology relationships within the clock tree.
\end{myproblem}

In this work, the key challenges are formulating unified double-side CTS design space and developing efficient multi-objective algorithm to explore Pareto-optimal solutions within this design space.

\section{Algorithms}
\label{sec:Algorithm}

\subsection{Overview}
The overall flow of our algorithm is shown in Fig. 4. It takes placement results (Placed DEF), PDK, and capacitance and connectivity constraints as input, and output a legal clock tree with buffers and nTSVs. The algorithm mainly consists of three steps: hierarchical clock routing, concurrent buffer and nTSV insertion, and skew refinement (SR). Based on this algorithm, we also support design space exploration of the double-side CTS solutions. We explain each step in details in the following sections.

\begin{figure}[tb]
    \centering
    \includegraphics[width=0.618\linewidth]{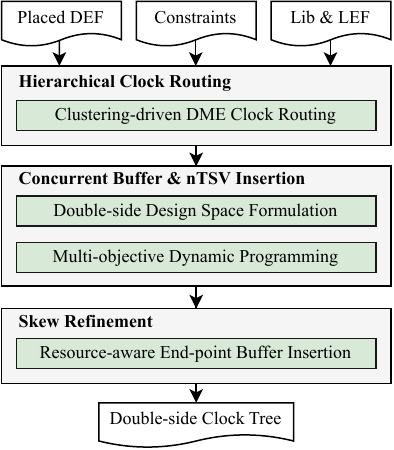}
    \caption{Overview of our algorithm framework.}
    \label{fig:flow}
    \vspace{-0.5cm}
\end{figure}

\subsection{Hierarchical Clock Routing}

In the modern CTS, the goal of clock routing is to firstly provide an initial clock tree topology that approximates the optimization of latency, skew, and wirelength. However, many follow-up timing-optimization stages, e.g., buffer insertion and sizing, make latency and skew more resilient to changes in topology, while the wirelength is still largely determined by the clock routing topology and impacts power significantly. We propose a hierarchical clock routing focuses on optimizing wirelength by combining clustering and deferred-merge-embedding (DME) clock routing.

\begin{figure}[tp]
    \centering
    \subfloat[High-level Clustering]{\includegraphics[width=0.31\columnwidth]{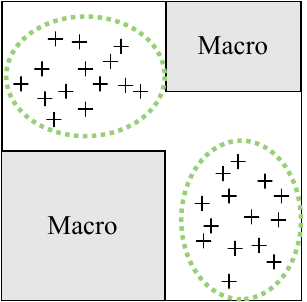}\label{fig:cluster-high}} \hspace{0.2cm}
    \subfloat[Low-level Clustering]{\includegraphics[width=0.31\columnwidth]{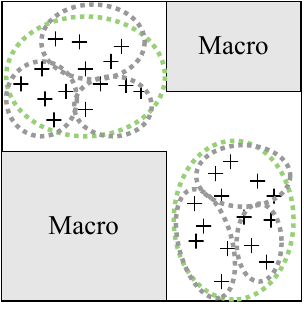}\label{fig:cluster-low}} \hfill\\
    \subfloat[Matching-based DME]{\includegraphics[width=0.31\columnwidth]{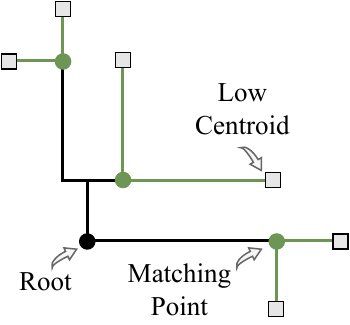}\label{fig:cluster-tree-matching}} \hspace{0.2cm}
    \subfloat[Hierarchical DME]{\includegraphics[width=0.31\columnwidth]{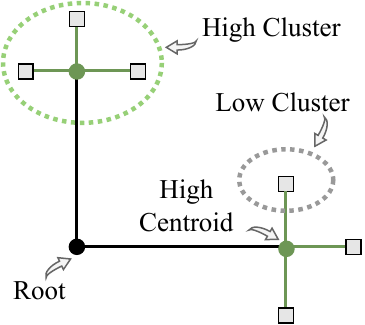}\label{fig:cluster-tree-clustering}} \hfill
    \caption{Dual-level clustering and DME-based clock routing. Symbol ``$\boldsymbol{+}$'' refers to sink.}
    \label{fig:dual-level-clustering}
\end{figure}


In our hierarchical clock routing, the clustering is performed at two sequential steps, i.e., high-level clustering and low-level clustering, as shown in \Cref{fig:cluster-high} and \Cref{fig:cluster-low}, respectively. High-level clustering groups the sinks into several large clusters with size $H_c$ by minimizing the total intra-cluster wirelength approximately. Low-level clustering then divides each large cluster into smaller ones with size $L_c$. The purpose of the dual-level clustering is to obtain a hierarchy according to the spatial proximity of sinks. We also record the centroids of both high- and low-level clustering solutions for later steps. K-means algorithm is adopted as the backbone for both clustering steps. In the experiments, we set $H_c$ to 3,000 and $L_c$ to 30 empirically. 

DME is widely used in many clock routing algorithms \cite{boese1992zero, edahiro1993clustering}. It helps to minimize skew and wirelength efficiently. A typical DME is based on matching, as shown in \Cref{fig:cluster-tree-matching}. However, such an approach is reported to have poor wirelength when dealing with imbalanced distribution of sinks. Therefore, with the clustering results, we perform DME clock routing with the low-level clustering centroids as leafs and the corresponding high-level clustering centroids as root, denoted as hierarchical DME as shown in \Cref{fig:cluster-tree-clustering}.

\subsection{Concurrent Buffer and nTSV Insertion}
In this section, we describe the details of double-side design space formulation and multi-objective DP.
 
\subsubsection{Double-side Design Space Formulation}
\begin{figure}[tb]
    \vspace{-0.3cm}
    \centering
    \includegraphics[width=0.76\columnwidth]{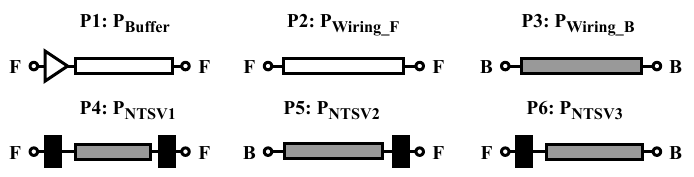}
    \caption{Patterns of wire segments labeled as P1 $\sim$ P6. \textbf{F} refers to the front side. \textbf{B} refers to the back side. The \textbf{right end} is close to sinks and the \textbf{left end} is close to clock root.}
    \label{fig:segments}
    \vspace{-0.5cm}
\end{figure}

The double-side design space is formulated by the discrete edge patterns and connectivity constraints. Different from the traditional buffer insertion on the single side, the edge patterns in our algorithm should adapt to double sides and the characteristics of buffers and nTSVs. We list six candidate edge patterns, denoted as pattern set $P$, in the generated clock tree from hierarchical clock routing, as shown in \Cref{fig:segments}. For instance, since the two pins of nTSVs are situated in different sides, resulting in side types of two endpoints of edge having distinct types as $\bm{\mathrm{P_{NTSV2}}}$ and $\bm{\mathrm{P_{NTSV3}}}$. However, the side types of two endpoints of $\bm{\mathrm{P_{NTSV1}}}$ are still \textbf{F} due to two nTSVs flipping side twice. Meanwhile, since the two pins of buffers are located in the front side, the side types of two endpoints of edge have to be \textbf{F} as $\bm{\mathrm{P_{Buffer}}}$. During buffers and nTSVs insertion process, the pattern decisions of adjacent edges cannot violate the \textbf{connectivity constraint}, that the shared vertex of any two edges in the clock tree must have the same side type.

\subsubsection{Multi-objective Dynamic Programming}

\begin{figure}[tb]
    \centering
    \includegraphics[width=0.9\linewidth]{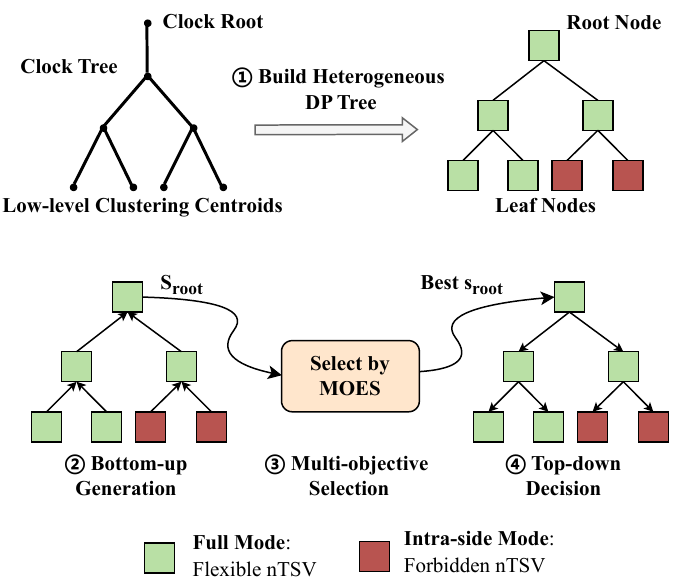}
    \caption{The process of building DP formulation form clock tree and the bottom-up and top-down with DP execution. Each edge corresponds to two sets marked by different colors.}
    \label{fig:dpflow}
    \vspace{-0.4cm}
\end{figure}

The multi-objective dynamic programming consists of four steps: build heterogeneous DP graph, bottom-up generation, multi-objective selection, and top-down decision, as shown in \Cref{fig:dpflow}. With these steps, we can concurrently insert buffers and nTSVs into the clock tree efficiently.

\textbf{Step 1 (Build Heterogeneous DP Tree):} To build the heterogeneous DP graph, we firstly represent each edge of the clock tree by a node in the DP graph. Two adjacent edges in the clock tree at different levels are connected by edge in DP graph as shown in \Cref{fig:dpflow}. Due to the structure of clock tree, the DP graph is formed as a tree rooted by the node corresponding to the edge of clock root. Then, the pattern selection, i.e., buffer and nTSV insertion, can be conducted on the DP tree. Notice that the clock tree and DP tree are both binary tree.

A novel idea of ours is to further configure the nodes in DP graph with two types of nTSVs inserting mode: full mode (flexible nTSV with $P_1\sim P_6$ allowed to select) and intra-side mode (forbidden nTSV with only $P_1\sim P_3$ allowed to select), which could be easily implemented by restricting the allowed patterns on the edges in our framework. Thus, by the inserting mode configurations, we could obtain a heterogeneous DP tree. To support the DP algorithm to explore the solutions in the DP tree, we utilize $S$ to represent the candidate solutions of each node and $s$ to represent the selected final solution of each node.

\textbf{Step 2 (Bottom-up Generation)}: In the process of bottom-up generation, we firstly set the undirected edges in DP tree by the topology of clock tree to directed edges, which all point to the root node finally. During the generation process, each node should undergo two operations: merging from two predecessor nodes and inserting in itself. We generated candidate solutions $S$ for all nodes starting from the leaf nodes. For leaf nodes without predecessor nodes, the edge end-point close to sinks are forced to front-side, which restricts the initial insertion of leaf nodes to $\{\mathrm{P_1, P_2, P_4, P_5}\}$ without merging. From the candidate leaf node solutions, we can generate the candidate solutions of successor nodes by merging each solution from one predecessor node with every solution from the other one. Meanwhile, these dependencies are recorded in merged solutions for fast traversal in Step 4. Notice that the merging operation is allowed which two predecessor solutions obeying the connectivity constraint. This rule can ensure we generate a legal double-side clock tree just by one turn of DP without any additional legalization steps, which improves the efficiency of our algorithm. 


After the merging operations, each node should conduct the inserting operations to assign patterns based on the merged solutions. The patterns must be selected from the set $P$, while the selection could be restricted by the specific inserting mode. For instance, if one node is configured to the intra-side mode, the patterns $\{\mathrm{P_4, P_5, P_6}\}$ involving nTSVs are forbidden to inserted in that node. By the delay model introduced in \Cref{sec:prelim-delay} and \cite{van1990buffer}, we could calculate the effective capacitance and path delay for solutions after inserting patterns. By the iterative execution, we finally obtain the candidate solutions at the root node, denoted as $S_{root}$ at Step 2 in \Cref{fig:dpflow}.

\textbf{Step 3 (Multi-objective Selection):} With the generated candidate solutions $S_{root}$, we try to select the final solution $s_{root}\in S_{root}$ considering multi-objective optimization. With the efficient computation structure of DP, we could easily record the latency, buffer count, and nTSV count for each node during the bottom-up generation.

Thus, we propose a multi-objective enhancement score to approach the multi-objective optimization. With the additional nTSV as resource for insertion, the candidate solutions at the root node have many more combinations of buffers, nTSVs, and different latencies. The distribution of candidate solutions is much more diverse than that in the single-side buffer insertion scenario, as observed in \ref{sec:res-moes} in the experiment. Therefore, we utilize the multi-objective enhancement score (MOES) to decide the final solution $s_{root}$ as follows.
\begin{equation}
    \mathrm{MOES} = \alpha l_{root} + \beta b_{root} + \gamma n_{root},
    \label{eq:score}
\end{equation}
where $l_{root}$, $b_{root}$, and $n_{root}$ are the values of latency, buffer count, and nTSV count for the candidate solutions at the root edge. $\alpha$, $\beta$, and $\gamma$ are manual parameters to weight each objective.

\textbf{Step 4 (Top-down Decision):} After the decision of $s_{root}$, we invert the direction of edges in the DP tree for top-down decision. By the recorded dependencies in the merged solutions at Step 2, we can quickly retrace the final solutions for all nodes.

\textbf{Pruning technique:} We extend the \textit{inferior solution} concept in \cite{van1990buffer}, that the effective capacitance and maximum path delay of one solution both worse than those of another solution means this solution will always be viewed as a bad candidate, to the double-side scenarios by pruning candidate solutions at front-side and back-side, respectively. This method ensures our DP algorithm is optimal in terms of latency. Meanwhile, to satisfy the constraints of the max-driven capacitance, we prune the solution with effective capacitance exceeding the maximum threshold.

\subsection{Skew Refinement}
\label{sec:alg-skew}

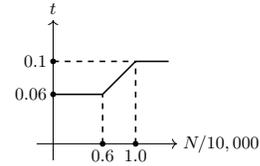
\begin{figure}
    \centering
    \resizebox{3.4cm}{!}{\begin{tikzpicture}[xscale=1.5, yscale=1.5]
        \draw[->] (-0.2,0) -- (1.5,0) node[right] {$N/10,000$};
        \draw[->] (0,-0.2) -- (0,1.5) node[above] {$t$};

        \draw[thick, black] (0,0.6) -- (0.6,0.6);
        \draw[thick, black, dashed] (0.6, 0) -- (0.6,0.6);
        \draw[thick, black] (0.6,0.6) -- (1,1);
        \draw[thick, black, dashed] (1, 0) -- (1,1);
        \draw[thick, black, dashed] (0, 1) -- (1,1);
        \draw[thick, black] (1, 1) -- (1.4,1);

        \fill (0.6,0) circle (1pt) node[below] {$0.6$};
        \fill (1.0,0) circle (1pt) node[below] {$1.0$};
        \fill (0,0.6) circle (1pt) node[left] {$0.06$};
        \fill (0,1) circle (1pt) node[left] {$0.1$};
    \end{tikzpicture}}
    \caption{Adaptive scale factor function $t\sim N/10,000$.}
    \label{fig:scale-t}
    \vspace{-0.6cm}
\end{figure}

As the DP in previous section mainly optimizes clock latency, we further introduce a resource-aware skew refinement technique to mitigate skew degradation by inserting buffers at end-points. This step is triggered when the skew is over $p$\% of the maximum latency. In the experiments, we set $p$ to 23. $N$ refers to the number of sinks. $t$ refers to an adaptive factor w.r.t. $N$, as shown in \Cref{fig:scale-t}.
\begin{enumerate}
    \item Set refined end-points number $n$ as $\min(N\times t, m)$. $m$ is the maximum number of refined end-points and set to 33 in the experiments.
    \item Refine $n$ end-points in descending order of delay by inserting one buffer at the low-level clustering centroids.
\end{enumerate}
By the observation that the skew and wire delay within lower clusters have a negligible impact on the overall path, this method is efficient and effective in mitigating skew.

\subsection{Design Space Exploration of Double-side CTS}
\label{sec:alg-dse}
\begin{figure}[tb]
    \centering
    \includegraphics[width=0.9\linewidth]{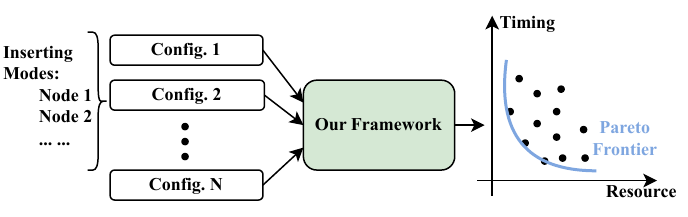}
    \caption{Design space exploration for double-side CTS.}
    \label{fig:dse}
    \vspace{-0.6cm}
\end{figure}

Based on the concurrent buffer and nTSV insertion approach, we further propose a general double-side design space exploration methodology that demonstrates superiority in multi-objective optimization of double-side CTS. The main idea is to control the inserting modes of nodes in DP tree, as shown in \Cref{fig:dpflow}. By setting up various configurations, users could explore more solutions in the objective space, as shown in \Cref{fig:dse}.

To make the DSE process easy to control, we allow users to control the inserting modes of nodes in DP tree by setting a fanout threshold. Nodes with fanout lower than the threshold will be configured as full mode, while those with fanout larger than the threshold will only allow intra-side mode. Furthermore, more sophisticated methods to control the inserting modes could be incorporated into our framework other than the simple heuristics. The concept of decoupling the DP execution flow and the controlling of nodes inserting allows users to avoid dealing with cumbersome details, e.g., timing calculation, but still have large multi-objective optimization space, which will promote more optimization techniques in double-side CTS problem.

\section{Experimental Results}
\label{sec:Results}
We perform the experiments on the Linux platform with a 20-core 2.40GHz Intel(R) Xeon(R) Silver 4210R CPU and 320GB memory. We take the ASAP7 PDK \cite{clark2016asap7} to perform our experiments 
and adopt the unit resistance and capacitance of back-side metal layers and nTSV from \cite{chen2021design}. These parameters are listed in \Cref{tab:tech}. We take designs from OpenROAD \cite{gitopenroad} and use its backend flow to generate benchmarks. The statistics of the benchmarks are listed in \Cref{tab:statistics}. Our framework is implemented using C++.

\subsection{Technology Settings}
We follow OpenROAD's convention to take the unit resistance and capacitance of M3 for the evaluation of delays in the front-side. For the back-side wires, we compute delays according to the actual usage of layers (\texttt{BM1}$\sim$\texttt{BM3}). We use the Elmore delay \cite{hu2015tau}, the slew model \cite{sitik2016design}, and the nonlinear delay model (NLDM) \cite{clark2016asap7} for delay computation. In our work, we follow the default flow in OpenROAD where one kind of buffer is used. This is a reasonable setting, because buffer sizing will be further optimized for skew minimization in the follow-up clock tree optimization after clock tree synthesis in a real design flow \cite{innovus, gitopenroad, ewetz2017clock}. We take \texttt{BUFx4\_ASAP7\_75t\_R} with a shape of $0.378nm \times 0.27nm$ as the buffer and \texttt{nTSV} with a shape of $0.27nm\times 0.27nm$, which is aligned to the other standard cells in the layout. The resistance and capacitance of one \texttt{nTSV} are $0.020k\Omega$ and $0.004fF$.

\begin{table}[tb!]
  \centering
  \caption{Layer resistances and capacitances \cite{chen2021design, clark2016asap7}.}
  \resizebox{0.7\linewidth}{!}{
  \begin{threeparttable}
    \begin{tabular}{|ccc|}
      \hline
      Layer & Unit Res. $(k\Omega/\mu m)$ & Unit Cap. $(f\mathrm{F}/\mu m)$ \\
      \hline \hline
      \texttt{M1} & 0.138890 & 0.11368 \\
      \texttt{M2} & 0.024222 & 0.13426 \\
      \texttt{M3} & 0.024222 & 0.12918 \\
      \texttt{M4} & 0.016778 & 0.11396 \\
      \texttt{M5} & 0.014677 & 0.13323 \\
      \texttt{M6} & 0.010371 & 0.11575 \\
      \texttt{M7} & 0.009672 & 0.13293 \\
      \texttt{M8} & 0.007431 & 0.11822 \\
      \texttt{M9} & 0.006874 & 0.13497 \\
      \hline
      \texttt{BM1$\sim$BM3} & 0.000384 & 0.116264 \\
      \hline
      \end{tabular} 
      \label{tab:tech}
    \end{threeparttable}
  }
\end{table}

\begin{table}[tp]
  \centering
    \caption{The statistics of benchmarks \cite{gitopenroad}.}
    \resizebox{0.7\linewidth}{!}{
    \begin{tabular}{|c|c|rrr|}
      \hline
      \multirow{2}[1]{*}{ID} & \multirow{2}[2]{*}{Design} & \multicolumn{3}{c|}{Statistics}  \\
       &  & $\#$Cells & $\#$FFs & Util. \\ \hline \hline
      C1&\texttt{jpeg}             & 54973  & 4380  & 0.50 \\
      C2&\texttt{swerv\_wrapper}   & 148407 & 14338 & 0.40 \\ 
      C3&\texttt{ethmac}           & 56851  & 10018 & 0.40 \\
      C4&\texttt{riscv32i}         & 11579  & 1056  & 0.50 \\ 
      C5&\texttt{aes}              & 29306 &  2072  & 0.50 \\
      \hline
      \end{tabular}
    }
  \label{tab:statistics}%
  \vspace{-0.3cm}
\end{table}%

\begin{table*}[tp]
  \vspace{-0.05in}
  \centering
    \caption{Comparison with recent studies on clock tree synthesis with back-side metal layers.}
    \resizebox{\textwidth}{!}{
    \begin{threeparttable}
    \begin{tabular}{|c|rrrr|rrrrrr|rrrrrr|}
    \hline
    \multirow{2}[2]{*}{Design} & \multicolumn{4}{c|}{\texttt{OpenROAD Buffered Clock Tree}$^\dagger$} & \multicolumn{6}{c|}{\texttt{OpenROAD Buffered Clock Tree} + \cite{veloso2023backside}} & \multicolumn{6}{c|}{\texttt{Ours}} \\
      & Latency & Skew & Buffers & nTSVs & Latency & Skew & Buffers & Clk WL & nTSVs & RT & Latency & Skew & Buffers & Clk WL & nTSVs & RT$^\ddagger$ \\
      & ($ps$) & ($ps$) &(\#)    & (\#)  & ($ps$) & ($ps$) & (\#)   & ($\times 10^6$) & (\#)  & ($s$) & ($ps$) & ($ps$) & (\#) & ($\times 10^6$) & (\#)  & ($s$) \\
    \hline \hline
    \texttt{C1}             & 246.154 & 37.189 & \textbf{167} & 0 & 172.027 & 40.171  & \textbf{167} & 4.768  & 189 & 4.351  & \textbf{77.694}  & \textbf{29.74}   & 172           & \textbf{3.664} & \textbf{130} & \textbf{0.285} \\
    \texttt{C2}   & 449.208 & 334.353 & 576 & 0 & 311.214 & 269.791 & 576 & 14.702 & 674 & 7.095          & \textbf{123.355} & \textbf{58.632}  & \textbf{571}  & \textbf{11.394}& \textbf{499} & \textbf{1.693} \\
    \texttt{C3}           & 214.629 & 25.953 & 375 & 0 & 159.982 & 20.929  & 375 & 6.019  & 487 & 4.470           & \textbf{90.229}  & \textbf{20.675}  & 375           & \textbf{5.326} & \textbf{256} & \textbf{1.149} \\
    \texttt{C4}         & 141.382 & 25.351 & 40  & 0 & 133.958 & 24.441  & 40  & 0.801 & 46 & 3.221              & \textbf{54.296}  & 20.231  & \textbf{33}   & \textbf{0.623}  & \textbf{24}  & \textbf{0.038} \\ 
    \texttt{C5}              & 213.993 & 75.261 & 79  & 0 & 192.928 & 78.307  & 79   & 1.723 & 89 & 3.436            & \textbf{90.829}  & \textbf{46.735}  & \textbf{74}   & \textbf{1.418} & 73  & \textbf{0.096} \\
    \hline
    \texttt{Ratio}            & 2.900 & 2.830 & 1.010 & - & 2.223 & 2.464 & 1.010 & 1.249 & 1.441 & 6.922 & \textbf{1.000} & \textbf{1.000} & \textbf{1.000} & \textbf{1.000} & \textbf{1.000} & \textbf{1.000} \\
    \hline
    \hline
    \multirow{2}[2]{*}{Design} & \multicolumn{4}{c|}{\texttt{Our Buffered Clock Tree}*} & \multicolumn{4}{c|}{\texttt{Our Buffered Clock Tree} + \cite{veloso2023backside}} & \multicolumn{4}{c|}{\texttt{Our Buffered Clock Tree} + \cite{bethur2023back}} & \multicolumn{4}{c|}{\texttt{Our Buffered Clock Tree} + \cite{bethur2024gnn}} \\
          & Latency & Skew & Buffers & nTSVs & Latency & Skew & Buffers & \multicolumn{1}{c|}{nTSVs} & Latency & \multicolumn{1}{c}{Skew} & Buffers & \multicolumn{1}{c|}{nTSVs} & Latency & Skew & Buffers & nTSVs  \\
          & ($ps$) & ($ps$) &(\#)    & (\#)  & ($ps$) & ($ps$) & (\#)   & \multicolumn{1}{c|}{(\#)} & ($ps$) & \multicolumn{1}{c}{($ps$)} & (\#) & \multicolumn{1}{c|}{(\#)} & ($ps$) & ($ps$) & (\#) & (\#) \\
    \hline \hline
    \texttt{C1}             & 144.603 & 41.839 & 189 & 0 & 130.420 & 54.757  & 189 & \multicolumn{1}{r|}{243} & 129.802 & \multicolumn{1}{r}{54.300}  & 189 & \multicolumn{1}{r|}{167} & 130.420 & 55.068  & 189 & 200 \\
    \texttt{C2}   & 273.704 & 70.390 & 588 & 0 & 244.554 & 117.383 & 588 & \multicolumn{1}{r|}{739} & 244.505 & \multicolumn{1}{c}{117.334} & 588 & \multicolumn{1}{r|}{551} & 244.554 & 117.383 & 588 & 576 \\
    \texttt{C3}           & 130.559 & 29.076 & \textbf{366} & 0 & 110.956 & 30.677  & \textbf{366} & \multicolumn{1}{r|}{418} & 113.196 & \multicolumn{1}{r}{32.917}  & \textbf{366} & \multicolumn{1}{r|}{385} & 110.956 & 30.677  & \textbf{366} & 382 \\
    \texttt{C4}         & 71.089  & \textbf{17.289} & 42  & 0 & 65.768  & 28.930  & 42  & \multicolumn{1}{r|}{50}  & 65.768  & \multicolumn{1}{r}{28.930}  & 42  & \multicolumn{1}{r|}{35}  & 65.768  & 28.930  & 42  & 47  \\ 
    \texttt{C5}              & 127.997 & 60.538 & 85  & 0 & 110.083 & 64.472  & 85  & \multicolumn{1}{r|}{109} & 110.083 & \multicolumn{1}{r}{63.693}  & 85  & \multicolumn{1}{r|}{72}  & 110.083 & 63.693  & 85  & \textbf{66}  \\
    \hline
    \texttt{Ratio}            & 1.714 & 1.245 & 1.037 &  -  & 1.516 & 1.683 & 1.037 & \multicolumn{1}{r|}{1.588} & 1.520 & \multicolumn{1}{r}{1.688}  & 1.037 & \multicolumn{1}{r|}{1.232} & 1.516 & 1.680 & 1.037 & 1.294 \\
    \hline
    \end{tabular}
    \begin{tablenotes}
      \item[$\dagger$] Use OpenROAD to generate buffered clock tree on front-side only. The Clk WL metric is the same as OpenROAD + \cite{veloso2023backside} since the same clock topology is used.
      \item[*] Use our framework to generate buffered clock tree on front-side only by three steps: conducting hierarchical clock routing, buffer insertion, and skew refinement. The Clk WL metric is the same as \texttt{Ours} since the same clock topology is used.
      \item[$\ddagger$] The runtime for other methods are omitted due to space limit, as their runtime is either similar to \texttt{Ours} or to \texttt{OpenROAD} + \cite{veloso2023backside}, according to which algorithm used to generate the buffered clock tree. 
    \end{tablenotes}  
  \end{threeparttable}
  }
  \label{tab:ours}%
  \vspace{-0.1in}
\end{table*}%

\subsection{Comparison with SOTA Methods}
We utilize the open-source tool OpenROAD \cite{gitopenroad} to evaluate the effectiveness of back-side metal layers and the benefits of our algorithm. We generate the \texttt{post-place} and \texttt{post-cts} DEF files \cite{chen2019datc}  from OpenROAD and use consistent evaluation methods and parameters for all designs. $\alpha$, $\beta$, and $\gamma$ are set to 1, 10, and 1. The fanout of \cite{bethur2023back} is set to 100 and the number of timing critical paths in \cite{bethur2024gnn} is set to 0.5. The inserting modes of all edges of our algorithm in \Cref{tab:ours} are set to full mode.

We also implement the method from \cite{veloso2023backside} as the baseline, which extremely optimizes latency. We take the \texttt{post-cts} DEF generated by the CTS tools in OpenROAD without resizing operations following CTS. According to \cite{veloso2023backside}, we flipped the nets above the low clustering centroids to the back-side by inserting nTSV as illustrated in \Cref{fig:clock-tree2} to minimize latency, denoted as \texttt{OpenROAD Buffered Clock Tree} + \cite{veloso2023backside} (\texttt{OpenROAD} + \cite{veloso2023backside} for short) in \Cref{tab:ours}. 

\Cref{tab:ours} summarizes the comparison between our framework and recent studies. 
Our algorithm outperforms the method \cite{veloso2023backside} in the clock latency by 2.223$\times$, skew by 2.464$\times$, number of buffers by 1.010$\times$, clock wirelength by 1.249$\times$, and number of nTSVs by 1.441$\times$. The significant reduction in latency and $\#$nTSVs comes from our hierarchical clock routing and concurrent buffer and nTSV insertion. We also compare with recent studies \cite{veloso2023backside,bethur2023back,bethur2024gnn} using the buffered clock tree generated by our framework, to demonstrate the effectiveness of the concurrent buffer and nTSV insertion. We can see that our algorithm could achieve better quality in almost all objectives. In addition, our algorithm is efficient. We achieve 6.992$\times$ speedup over \texttt{OpenROAD} + \cite{veloso2023backside}. 

\subsection{Effectiveness of MOES}
\label{sec:res-moes}
\begin{figure}[tp]
  \vspace{-0.2cm}
  \centering
  \includegraphics[width=0.9\columnwidth]{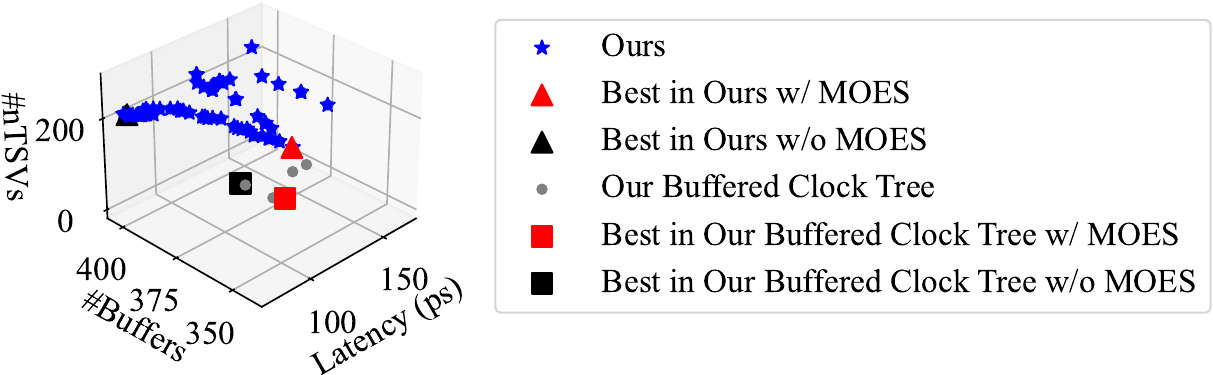}
  \caption{The effectiveness of MOES with \texttt{C3 (ethmac)} under \texttt{Ours} and \texttt{Our Buffered Clock Tree}.}
  \label{fig:moes}
  \vspace{-0.1in}
\end{figure}

In \Cref{fig:moes}, we validate the effectiveness of MOES (\Cref{sec:Algorithm}) in concurrent buffer and nTSV insertion compared with the solo buffered clock tree. The points show the best results achieved by using MOES and minimal latency deviate far away in the double-side scenario (see two triangles), while they are much closer in the single-side scenario (see two squares). The reason is that the double-side scenario enlarges the design space, where more combinations of buffers and nTSVs will be preserved at the end of DP and should be carefully considered. Therefore, an objective function considering holistic factors, like MOES, is necessary to improve the solution quality for double-side CTS.

\subsection{Effectiveness of Skew Refinement}
\begin{figure}[tp]
  \centering
  \includegraphics[width=1\columnwidth]{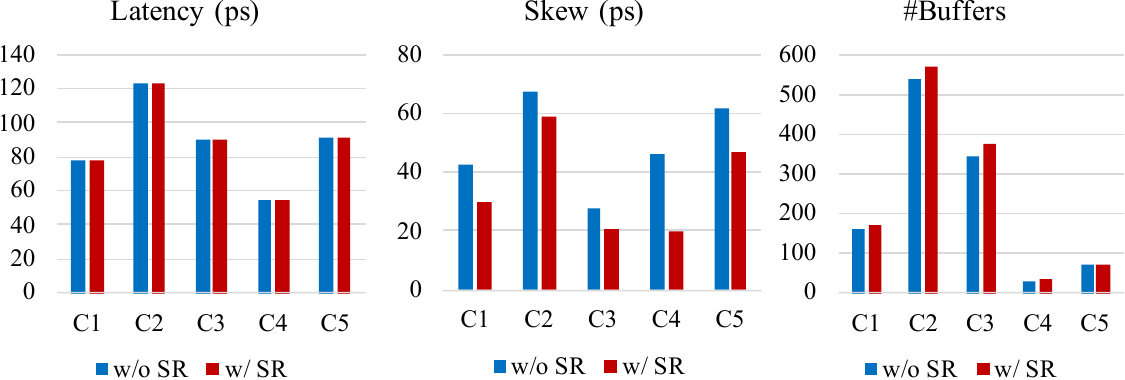}
  \caption{Effectiveness of skew refinement. 
  }
  \label{fig:skew-refine}
  \vspace{-0.5cm}
\end{figure}

In \Cref{fig:skew-refine}, we demonstrate the effectiveness of skew refinement by the designs in \Cref{tab:statistics}. In the skew refinement, we utilize the method in Section \ref{sec:alg-skew} to pick up the paths that need skew refinement. 
From \Cref{fig:skew-refine}, the skew can be reduced significantly, while the increases in latency and \#buffers are very ignorable. This indicates that the method can effectively reduce skew as a post-processing step, which can also provide a better initial solution for the follow-up clock optimization stages.

\subsection{Comparison on Design Space Exploration}
To further verify the superiority of our systematic double-side framework, we perform design space exploration on different methods. We set the fanout ranging from 20 to 1000 with step 10 in our DSE flow as introduced in \Cref{sec:alg-dse}. Meanwhile, we set the fanout of \cite{bethur2023back} also ranging from 20 to 1000 with step 10 and the percentage of critical paths of \cite{bethur2024gnn} ranging from 0.2 to 0.9 with step 0.05. To make the comparison fair, the buffered clock tree is all generated by our algorithm. 

\begin{figure}[tp]
  \vspace{-0.2cm}
  \centering
  \includegraphics[width=0.9\columnwidth]{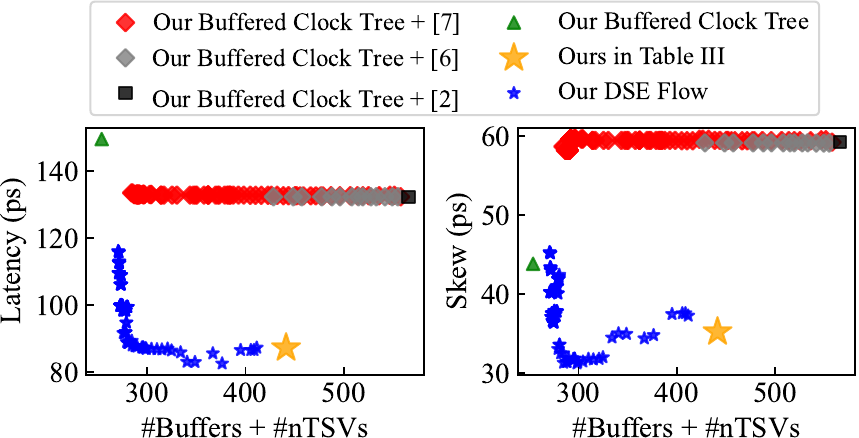} \hfill
  \caption{The comparison of latency and skew of different flows.}
  \label{fig:pareto}
  \vspace{-0.4cm}
\end{figure}

In \Cref{fig:pareto}, we compare our DSE flow with recent methods \cite{veloso2023backside, bethur2023back, bethur2024gnn} in exploration of multi-objective solutions. 
We can see that the solution space of these methods are restricted to the buffered clock trees, and cannot effectively explore better latency or skew even given more nTSVs. 
A possible reason is the target of buffered clock tree is to minimize clock latency by inserting buffers to shorten wirelength, while the ability of nTSV insertion will be weakened with the wirelength decreasing. 
Instead, our DSE flow controls the inserting modes in concurrent buffer and nTSV insertion allows much larger solution space. By simply sweeping the fanout threshold, we can find the Pareto frontier trading off latency, skew, buffers, and nTSVs. As our algorithm is very efficient, users can search for suitable solutions with our DSE flow at low cost. 


%

\section{Conclusion}
\label{sec:Conclusion}

In this work, we propose a systematic framework for clock tree synthesis with double-side metal layers and nTSVs.
We propose hierarchical clock routing, concurrent buffer and nTSV insertion, and skew refinement to optimize clock tree in multiple objectives.
Compared with the recent method \cite{veloso2023backside} with extreme optimization on latency, we can optimize the clock latency by 2.223$\times$, skew by 2.464$\times$, number of buffers by 1.010$\times$, clock wirelength by 1.249$\times$, and number of nTSVs by 1.441$\times$, respectively, with 6.922$\times$ speed-up.
Meanwhile, our framework offers a DSE flow for multi-objective exploration to further boost the technology and design progressing.
In the future, we will investigate placement and routing effects on double-side CTS and develop methodologies for full-flow optimization.


\vspace{-0.06in}
\section*{Acknowledgements}
The project is supported in part by
Grant QYJS-2023-2303-B,
the Natural Science Foundation of Beijing, China (Grant No.~Z230002), 
the 111 project (B18001),
and Research Grants Council of Hong Kong SAR (No.~RFS2425-4S02 and No.~CUHK14211824).

{
\small
\bibliographystyle{IEEEtran}
\bibliography{dac25}
}

\end{document}